\documentclass[twocolumn,aps,prl,showpacs,superscriptaddress]{revtex4}

\usepackage{graphicx}
\usepackage{amsmath,amssymb}
\usepackage{color}

\newcommand{\beq}{\begin{equation}}
\newcommand{\eeq}{\end{equation}}
\newcommand{\beqa}{\begin{eqnarray}}
\newcommand{\eeqa}{\end{eqnarray}}
\newcommand{\ket} [1] {\vert #1 \rangle}
\newcommand{\bra} [1] {\langle #1 \vert}
\newcommand{\braket}[2]{\langle #1 | #2 \rangle}
\newcommand{\proj}[1]{\ket{#1}\!\bra{#1}}
\newcommand{\mean}[1]{\langle #1 \rangle}
\newcommand{\identity}{\mbox{$1 \hspace{-1.0mm} {\bf l}$}}

\begin{document}


\title{Two Fundamental Experimental Tests of Nonclassicality with Qutrits}



\author{Johan Ahrens}
 \affiliation{Department of Physics, Stockholm University, S-10691,
 Stockholm, Sweden}

\author{Elias Amselem}
 \affiliation{Department of Physics, Stockholm University, S-10691,
 Stockholm, Sweden}

\author{Ad\'an Cabello}
 \affiliation{Departamento de F\'{\i}sica Aplicada II, Universidad de
 Sevilla, E-41012 Sevilla, Spain}
 \affiliation{Department of Physics, Stockholm University, S-10691,
 Stockholm, Sweden}

\author{Mohamed Bourennane}
 \affiliation{Department of Physics, Stockholm University, S-10691,
 Stockholm, Sweden}


\date{\today}



\begin{abstract}
We report two fundamental experiments on three-level quantum systems (qutrits). The first one tests the simplest task for which quantum mechanics provides an advantage with respect to classical physics. The quantum advantage is certified by the violation of Wright's inequality, the simplest classical inequality violated by quantum mechanics. In the second experiment, we obtain contextual correlations by sequentially measuring pairs of compatible observables on a qutrit, and show the violation of Klyachko \emph{et al.}'s inequality, the most fundamental noncontextuality inequality violated by qutrits. Our experiment tests exactly Klyachko {\it et al.}'s inequality, uses the same measurement procedure for each observable in every context, and shows that the violation does not depend on the order of the measurements.
\end{abstract}



\maketitle


{\em Introduction.---}In classical physics, there is no contradiction in considering that systems like balls and coins have preestablished properties like
position and velocity that are independent of whether one actually measures them or not. However, according to quantum mechanics, the results of experiments on systems such as atoms and photons do not correspond to preestablished properties. A natural and fundamental question is: Which is the simplest quantum system in which this difference between classical and quantum physics can be observed?

For instance, this difference is shown in the violation of Bell inequalities \cite{ADR82,WJSWZ98,MMMOM08}, which requires at least two measurements on, at least, two separate subsystems, thus requiring a physical system of dimension four (i.e., with four perfectly distinguishable states). However, even before Bell inequalities were discovered, Kochen and Specker \cite{Specker60,KS67}, and Bell \cite{Bell66} pointed out that the classical/quantum conflict occurs even in a simpler three-level quantum system, or qutrit (e.g., a spin-1 particle). This implies that entanglement is not needed for the generation of nonclassical correlations, since, by definition, a single qutrit cannot be in an entangled state. Indeed, by performing a sequence of compatible \cite{Peres93} measurements on the same single qutrit, contextual correlations can be obtained.

The classical/quantum conflict can be revealed even without correlations. The simplest test producing classically impossible results was first described by Wright \cite{Wright78} and is based on the probabilities of the results of five single measurements on a qutrit. Later on, Klyachko \emph{et al.} derived a noncontextuality inequality (i.e., satisfied by all theories in which the measurement results are independent of any compatible measurement) for qutrits using correlations. Klyachko \emph{et al.}'s inequality is the simplest noncontextuality inequality violated by quantum mechanics, in the sense that there is no conflict for systems of lower dimension or with inequalities with fewer terms.

It should be noted that a recent experiment on qutrits \cite{LLSLRWZ11} does not test Klyachko \emph{et al.}'s inequality, but an inequality with extra correlations. In \cite{LLSLRWZ11} it is left as an ``open question'' whether any experimental apparatus can be designed to test exactly Klyachko \emph{et al.}'s inequality.

In this Letter we report experimental violations of both Wright's and exactly Klyachko \emph{et al.}'s inequality with qutrits. We first introduce both inequalities and show that they are actually connected. Then, we describe the experimental setups corresponding to each test. Finally, we present the experimental results and discuss them.

{\em Wright's inequality.---}Wright's inequality \cite{Wright78} is the simplest classical inequality violated by quantum mechanics. It can be proven that the simplest set of questions such that the {\em sum} of the probabilities of obtaining a yes answer is higher in quantum mechanics than in classical physics is precisely the one in which 5 questions $Q_i$ are such that $Q_i$ and $Q_{i+1}$ (with the sum modulo 5) are exclusive. In each run of the game, one of these questions is picked at random. The goal of the game is to obtain as many yes answers as possible.

Below we give an example that illustrates an optimal classical strategy. The system is prepared in a random mixture of 5 classical states $0$, $1$, $2$, $3$, and $4$, and
the player provides the following 5 yes-no questions:
 $Q_0=$``0 or 1?'' (denoting the question ``Is the system in one of the states $0$ or $1$?''),
 $Q_1=$``2 or 3?'',
 $Q_2=$``0 or 4?'',
 $Q_3=$``1 or 2?'', and
 $Q_4=$``3 or 4?''.
If $P(+1|Q_{i})$ denotes the probability of obtaining a yes answer when $Q_i$ is asked, then, in our example, $P(+1|Q_{i})=\frac{2}{5}$ for any $i=0,\ldots,4$. Therefore, $\sum_{i=0}^4 P(+1|Q_{i})=2$. It can be proven that no other preparation or set of classical questions provides a better solution; for any set of 5 classical questions with the previous exclusiveness constraints, the following inequality holds:
\begin{equation}
 {\cal W}:= \sum_{i=0}^4 P(+1|Q_{i}) \le 2.
 \label{Wrightinequality}
\end{equation}
The upper bound follows from the fact that the maximum number of questions $Q_{i}$ that simultaneously can have a yes answer is $2$. Inequality \eqref{Wrightinequality} is Wright's inequality.

According to quantum mechanics, the system can be in a superposition of states and the questions can refer to such a superposition. This provides an advantage. For example, the quantum system is prepared in the qutrit state
\begin{equation}
 \bra{\psi}=(0,0,1)
 \label{state}
\end{equation}
and the questions correspond to the 5 operators $Q_i=2\proj{v_{i}}-\identity$, which in quantum mechanics represent 5 observables with possible outcomes $+1$ and $-1$ (corresponding to yes and no, respectively). $Q_i$ and $Q_{i+1}$ are compatible and exclusive. Correspondingly, the 5 projectors $\proj{v_{i}}$ are such that $\proj{v_{i}}$ and $\proj{v_{i+1}}$ are orthogonal. Specifically, to obtain the maximum quantum value, we chose
\begin{subequations}
 \label{vectors}
\begin{align}
 \bra{v_{0}} &= N\left(1,0,r\right), \\
 \bra{v_{1,4}} &= N\left(c,\pm s,r\right),\\
 \bra{v_{2,3}} &= N\left(C,\mp S,r\right),
\end{align}
\end{subequations}
with $r=\sqrt{\cos{\left(\frac{\pi}{5}\right)}}$,
$c=\cos{\left(\frac{4\pi}{5}\right)}$,
$s=\sin{\left(\frac{4\pi}{5}\right)}$,
$C=\cos{\left(\frac{2\pi}{5}\right)}$,
$S=\sin{\left(\frac{2\pi}{5}\right)}$, $N=1/ \sqrt{1+r^2}$, and $\identity$ is the identity matrix.
Then, quantum mechanics predicts that
\begin{equation}
 {\cal W}_{\rm QM} = \sum_{i=0}^4 |\braket{v_i}{\psi}|^2=\sqrt{5}
 \approx 2.236,
\end{equation}
which violates Wright's inequality (\ref{Wrightinequality}).

It can be proven that the simplest set of questions (such that any of them belongs to an exclusive pair) and constraints for which quantum mechanics provides an advantage is precisely the one in which 5 questions $Q_i$ are such that $Q_i$ and $Q_{i+1}$ (with the sum modulo 5) are exclusive.
It can also be proven that the simplest system with quantum advantage has $d=3$. Finally, it can be proven that the maximum quantum violation (for any $d$) is precisely $\sqrt{5}$ (see Supplemental Material).


{\em Klyachko {\it et al.}'s noncontextuality inequality.---}The simplest experiment showing contextual correlations in a qutrit is the violation of Klyachko {\it et al.}'s inequality \cite{KCBS08}. This inequality defines the only nontrivial facet of the polytope of classical (noncontextual) correlations, and completely separates noncontextual from contextual correlations \cite{KCBS08}. The importance of observing contextual correlations between the results of sequential measurements on the same qutrit comes from the fact that these correlations cannot be attributed to entanglement, since, by the definition of entanglement, a single qutrit cannot be in an entangled state. This is not the case in recent experiments showing quantum contextual correlations on two-qubit systems \cite{KZGKGCBR09,BKSSCRH09,ARBC09,MRCL10}.

To put Klyachko {\it et al.}'s inequality in the frame of the classical game introduced before, we now ask two questions $Q_{i}$ and $Q_{i+1}$, one immediately after the other. By collecting all answers (yes, yes), (yes, no), (no, yes), and (no, no), we can calculate the average of obtaining the same results (i.e., $Q_{i}Q_{i+1}=+1$) or different results (i.e., $Q_{i}Q_{i+1}=-1$). For our classical strategy, for each pair of exclusive questions $Q_{i}$ and $Q_{i+1}$, we obtain $\mean{ Q_{i}Q_{i+1}}=-\frac{3}{5}$. In other words, when asking $Q_{i}$ and $Q_{i+1}$ sequentially, on average, 4 out of 5 times we obtain different answers, and 1 out of 5 times we obtain equal answers. If we now sum over all possible pairs $Q_{i}$, $Q_{i+1}$, we obtain $-3$, which can be proven to be the classical lower bound for the sum. Therefore
\begin{equation}
 \kappa := \sum_{i=0}^{4}\mean{Q_{i}Q_{i+1}} \geq -3,
 \label{KCBS}
\end{equation}
which is Klyachko {\it et al.}'s inequality \cite{KCBS08}.

The maximum quantum violation of Klyachko {\it et al.}'s inequality \eqref{KCBS} is attained for the same qutrit state $\ket{\psi}$ and observables $Q_{i}$ providing the maximum quantum violation of Wright's inequality \eqref{Wrightinequality}.
The maximum quantum violation (for any $d$) of inequality \eqref{KCBS} is
\begin{equation}
 \kappa_{\mathrm{QM}} = 5- 4\sqrt{5} \approx -3.944.
\end{equation}

We would like to note that {\L}apkiewicz \emph{et al.} \cite{LLSLRWZ11} have recently performed an experiment aiming to test a noncontextuality inequality with 6 measurements and 6 correlations (4 of them contained in Klyachko {\em et al.}'s inequality, plus two more which, added up, act as substitutes for the fifth correlation in Klyachko {\em et al.}'s inequality). However, the 6-correlation inequality tested by {\L}apkiewicz {\em et al.} is not a facet of the simplest polytope of noncontextual correlations (i.e., it does not belong to the simplest set of inequalities that separates noncontextual from contextual correlations). Moreover, {\L}apkiewicz {\em et al.}'s experiment cannot be considered a proper test of a noncontextuality inequality, since the same observable is measured with different setups in different contexts (see the Supplemental Material for further discussion).


\begin{figure}[t]
\centerline{\includegraphics[width=0.82\columnwidth]{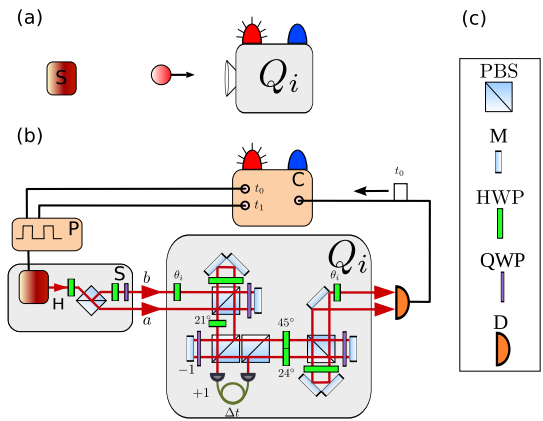}}
\caption{Experimental setup for experiment 1. (a) Scheme for single projective measurement $Q_i$. The red and blue lamps correspond to unsuccessful (no) and successful (yes) projection, respectively. (b) Setup for creating a qutrit and performing $Q_i$. A pulse generator, P in the figure, is trigging the attenuated diode laser in the source, S in the figure. The setup consists of a source of horizontally polarized single photons followed by a half wave plate (HWP) and a polarizing beam splitter (PBS), allowing us to prepare any probability distribution of a photon in modes $a$ and $b$. The orientation of the HWP in mode $b$ sets the polarization state of that mode. The output of the source is connected to the input of operator $Q_{i}$, which has detectors in its output. (c) Symbol definition of the optical elements used in the setup: polarizing beam splitter (PBS), mirror (M), half wave plate (HWP), quarter wave plate (QWP), and single photon detector (D).}\label{setup1}
\end{figure}

\begin{figure}[tb]
\centerline{\includegraphics[width=1\columnwidth]{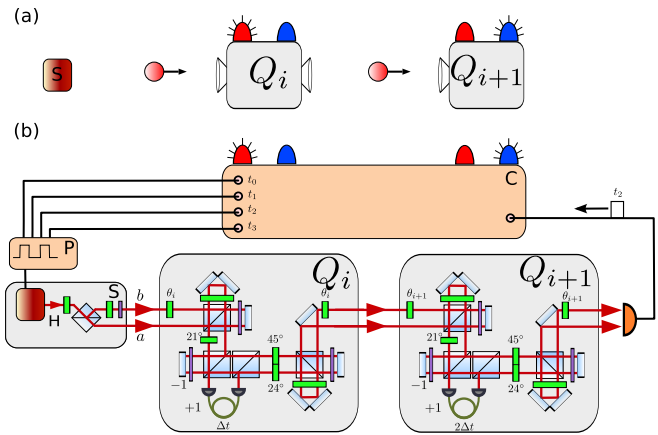}}
\caption{Experimental setup for experiment 2. (a) Scheme for the sequential measurement on pairwise compatible observables $Q_i$ and $Q_{i+1}$. The red and blue lamps correspond to the eigenvalues $-1$ (no) and $+1$ (yes), respectively. (b) Setup for performing the sequential measurements $Q_i$ and $Q_{i+1}$. The device for $Q_i$ is exactly the same as described in Fig.~\ref{setup1}; the device for the second measurement $Q_{i+1}$ is also the same, except for a longer time delay of $2 \Delta t$. A click at the detectors at the time slots $t_0$, $t_1$, $t_2$, and $t_3$ corresponds to, respectively, the answer (no, no), (yes, no), (no, yes), and (yes, yes). The preparation of the qutrit state and the symbols are the same as in Fig.~\ref{setup1}.}
\label{setup2}
\end{figure}

{\em Experimental setup.---}Experiment 1 (see Fig.~\ref{setup1}) tests the quantum violation of inequality (\ref{Wrightinequality}) and experiment 2 (see Fig.~\ref{setup2}) tests the quantum violation of inequality (\ref{KCBS}). In both experiments, the qutrit is defined by means of the polarization and path degrees of freedom of a single photon. The two spatial modes are labeled $a$ and $b$ and, by design, the polarization in mode $a$ is enforced to be horizontal (see Fig. 1). The encoding is
\begin{equation}
 \ket{0}=\ket{H,b},\;\;\;\;\ket{1}=\ket{V,b},\;\;\;\;\ket{2}=\ket{H,a},
\end{equation}
where $H$ denotes horizontal polarization and $V$ denotes vertical polarization. To ensure that the system stays a qutrit, we make sure that a potential $\ket{V,a}$ component is always associated to loss and thus never expands the Hilbert space.


{\bf Experiment 1.} The setup in Fig.~\ref{setup1}(b) allows for the preparation of all required states and the projections on the eigenstates of all required operators. The questions $Q_i$ are implemented through a time multiplexing scheme. A yes answer to $Q_i$ corresponds to a successful projection and is indicated by the arrival time $t_{1}$ of the photon (or, equivalently, by a blue lamp flashing; see Fig.~\ref{setup1}), the no answer corresponds to an unsuccessful projection and is indicated by the arrival time $t_{0}$ of the photon (or a red lamp flashing; see Fig.~\ref{setup1}).

The goal of the measurement is to distinguish the eigenstates corresponding to different eigenvalues of $Q_i$. Any qutrit pure state can be expressed as $\alpha \ket{H,b} + \beta \ket{V,b} + \gamma \ket{H,a}$, where $\alpha$, $\beta$, and $\gamma$ are complex numbers. To implement $Q_i$ we prepare the eigenstate $\ket{v_i}$, which corresponds to the positive eigenvalue. The polarization in mode $b$ will first be rotated to obtain the state $\beta' \ket{V,b} + \gamma \ket{H,a}$ (this can always be done, since the rotation of the polarization is performed only in mode $b$). Then, the part of the state in mode $a$ is transferred into mode $b$; this process also flips the polarization. This leads to the state $\beta' \ket{H,b} + \gamma \ket{V,b}$. The state can now be rotated to $\ket{H,b}$, which is coupled to a delay line adding a delay of $\Delta t=50$\,ns so that it can be distinguished from the orthogonal states by the time slots $t_{1}$ and $t_{0}$. All of the previous steps (except for the time delay) are then performed in reverse order to reprepare the eigenstate of the observable for further processing. A useful property of our implementation of the 5 operators $Q_i$ is that they are exactly the same, up to a half wave plate rotation; see Fig.~\ref{setup1}(b). To ask the question $Q_i$, one rotates both the first and last half wave plate by an angle $\theta_i = $ $45^\circ$, $117^\circ$, $9^\circ$, $81^\circ$, and $153^\circ$ for $i=0, 1, 2, 3$, and $4$, respectively.


{\bf Experiment 2.} When we perform two sequential measurements corresponding to pairwise compatible observables $Q_i$ and $Q_{i\pm 1}$, the first measurement is exactly the same one as described above. The eigenstates of observable $Q_i$ with eigenvalue $+1$ or $-1$ are again mapped to different time slots using the same time multiplexing detection method. The setup for the second measurement is the same as that for the first, except for the delay line, which is twice as long, i.e., $2 \Delta t$ see Fig.~\ref{setup2}(b). After passing the devices for $Q_i$ and $Q_{i+1}$, the photon can be registered by a detector at four equally distributed time slots. These time slots are $t_0$, $t_1 = t_{0} + \Delta t$, $t_2 = t_{0} + 2 \Delta t$, and $t_3 = t_{0} + 3 \Delta t$, and correspond, respectively, to the answers (no, no), (yes, no), (no, yes), and (yes, yes).

Another distinguishing feature of this approach is that it allows both the measurement of $Q_i$ followed by $Q_{i+1}$, and also the measurement of $Q_{i+1}$ followed by $Q_i$. This allows us to test each of the correlations in inequality (\ref{KCBS}) in every possible order.


In both experiments, the single-photon source was a pulsed diode laser emitting at $780$\,nm with a pulse width of $3$\,ns and a repetition rate of $100$\,kHz; see Fig.~\ref{setup1}(b) and Fig.~\ref{setup2}(b). The laser was attenuated so that the two-photon coincidences were negligible. The visibility ranges achieved were between $80\%$ and $90\%$ for each $Q_i$. A single-photon detector was placed in each output mode $a$ and $b$. All detector signals and timing trigger signals (which define the measurement time slots) were registered using a multichannel coincidence logic with a time window of $1.7$\,ns.


{\em Experimental results and discussions.---}The experimental results for experiment 1 testing inequality (\ref{Wrightinequality}) are shown in Table~\ref{TableI}. We observe a clear violation of inequality (\ref{Wrightinequality}). It can be noted that the obtained experimental value is slightly higher than the maximum quantum value. This can be explained by the fact that the latter is obtained under the assumption that the measured questions/operators are perfectly exclusive, while perfect exclusiveness is difficult to guarantee experimentally. In our implementation, a perfect question $Q_i$ would need an interference visibility of $100\%$ to guarantee a perfect exclusiveness with $Q_{i+1}$ (see Supplemental Material for more details).


\begin{table}[h]
\caption{\label{TableI}Experimental results for the violation of inequality ${\cal W} \le 2$, the theoretical quantum bound for an ideal experiment is ${\cal W}_{\rm QM} =\sqrt{5} \approx 2.236$.}
\begin{ruledtabular}
{
\begin{tabular}{c c}
$P(+1|Q_{0})$ & $0.4600 \pm 0.012$ \\
$P(+1|Q_{1})$ & $0.4544 \pm 0.012$ \\
$P(+1|Q_{2})$ & $0.4603 \pm 0.016$ \\
$P(+1|Q_{3})$ & $0.4610 \pm 0.011$ \\
$P(+1|Q_{4})$ & $0.4566 \pm 0.010$ \\
${\cal W}$    & $2.292  \pm 0.028$ \\
\end{tabular}
}
\end{ruledtabular}
\end{table}


The experimental results for experiment 2 testing inequality (\ref{KCBS}) are shown in Table~\ref{TableII}. The measurements are performed in all possible orders to show that the violation does not depend on the order. The experimental results in Table \ref{TableII} show a clear violation of inequality (\ref{KCBS}), in good agreement with the quantum mechanics prediction.


\begin{table}[h]
\caption{\label{TableII}Experimental results for the violation of inequality $\kappa \leq 3$, the theoretical quantum bound for an ideal experiment is $\kappa_{\mathrm{QM}} = 5- 4\sqrt{5} \approx -3.944.$}
\begin{ruledtabular}
{
\begin{tabular}{c c c c}
$\langle Q_{0}Q_{1}\rangle$ & $-0.712 \pm 0.002$ & $\langle Q_{1}Q_{0}\rangle$ & $-0.785 \pm 0.003$ \\
$\langle Q_{1}Q_{2}\rangle$ & $-0.706 \pm 0.002$ & $\langle Q_{2}Q_{1}\rangle$ & $-0.781 \pm 0.003$ \\
$\langle Q_{2}Q_{3}\rangle$ & $-0.704 \pm 0.002$ & $\langle Q_{3}Q_{2}\rangle$ & $-0.774 \pm 0.003$ \\
$\langle Q_{3}Q_{4}\rangle$ & $-0.708 \pm 0.002$ & $\langle Q_{4}Q_{3}\rangle$ & $-0.774 \pm 0.003$ \\
$\langle Q_{4}Q_{0}\rangle$ & $-0.706 \pm 0.002$ & $\langle Q_{0}Q_{4}\rangle$ & $-0.782 \pm 0.003$ \\
$\kappa$                    & $-3.536 \pm 0.005$ & $\kappa$                    & $-3.896 \pm 0.006$ \\
\end{tabular}
}
\end{ruledtabular}
\end{table}


The main source of systematic error was due to the optical interferometers involved in the measurements, the imperfect overlapping and coupling of the light modes, and the polarization components. Errors were inferred from propagated Poissonian counting statistics of the raw detection events and from 10 measurement samples to capture the error due to the drift over the measurement time. The number of detected photons was approximately $3\times10^4$ per second and the total measurement time for each of the 5 pairs of observables was $1$\,s for each run. Our measurement procedure was to first calibrate the interferometers and then start the measurement where all terms in (\ref{KCBS}) are measured one after the other. In Table~\ref{TableII}, the second column has higher values than the first, since it was impossible to replicate exactly the calibration of the interferometers between the two runs.


{\em Conclusions.---}We have reported two experiments on qutrits showing nonclassical properties. Unlike previous experiments on pairs of qubits \cite{KZGKGCBR09,BKSSCRH09,ARBC09,MRCL10}, here the nonclassical properties cannot be attributed to entanglement.

Experiment 1 tests Wright's inequality \cite{Wright78}, which is the simplest inequality based on outcome probabilities of 5 exclusive yes-no questions, that provides a better-than-classical solution. Experiment 2 tests Klyachko \emph{et al.}'s inequality \cite{KCBS08}, which is the most fundamental correlation inequality satisfied by noncontextual theories and violated by single qutrits. Unlike {\L}apkiewicz \emph{et al.}'s experiment \cite{LLSLRWZ11}, our experiment 2 tests exactly Klyachko \emph{et al.}'s inequality with 5 observables and 5 correlations, and tests the correlations in any possible order. Unlike previous experiments with sequential measurements on photons, in our experiment 2, the results of the sequential measurements are encoded in different time slots, avoiding much more complicated alternatives \cite{ARBC09}, and allowing us to show that the violation of the inequality does not depend on the order of the measurements. Both experiments are of fundamental importance to understand the difference between quantum and classical physics in situations where entanglement is not present, and open the door to new applications in quantum information processing based on simple quantum systems.


\begin{acknowledgments}
The authors thank I.\ Bengtsson, A.\ Klyachko, J.-\AA.\ Larsson, R.W.\ Spekkens, and K.\ Svozil for discussions. This work was supported by the Swedish Research Council (VR), the Linnaeus Center of Excellence ADOPT, the Spanish Projects Nos.\ FIS2008-05596 and FIS2011-29400, and the Wenner-Gren Foundation.
\end{acknowledgments}



\end{document}